# Artificial Intelligence-driven Image Analysis of Bacterial Cells and Biofilms

Shankarachary Ragi, Md Hafizur Rahman, Jamison Duckworth, Kalimuthu Jawaharraj, Parvathi Chundi, and Venkataramana Gadhamshetty

**Abstract**— The current study explores an artificial intelligence framework for measuring the structural features from microscopy images of the bacterial biofilms. *Desulfovibrio alaskensis* G20 (DA-G20) grown on mild steel surfaces is used as a model for sulfate reducing bacteria that are implicated in microbiologically influenced corrosion problems. Our goal is to automate the process of extracting the geometrical properties of the DA-G20 cells from the scanning electron microscopy (SEM) images, which is otherwise a laborious and costly process. These geometric properties are a biofilm phenotype that allow us to understand how the biofilm structurally adapts to the surface properties of the underlying metals, which can lead to better corrosion prevention solutions. We adapt two deep learning models: (a) a deep convolutional neural network (DCNN) model to achieve semantic segmentation of the cells, (d) a mask region-convolutional neural network (Mask R-CNN) model to achieve instance segmentation of the cells. These models are then integrated with moment invariants approach to measure the geometric characteristics of the segmented cells. Our numerical studies confirm that the Mask-RCNN and DCNN methods are 227x and 70x faster respectively, compared to the traditional method of manual identification and measurement of the cell geometric properties by the domain experts.

—————————— ◆ ——————————

## 1 INTRODUCTION

SULFATE-reducing bacteria (SRB) are widely implicated in microbiologically influenced corrosion (MIC) of metals, costing billions of dollars annually. Understanding the phenotypical growth characteristics of biofilms over time, particularly the size and shape of the cells in biofilms on the metal surfaces and how they adapt in corrosive environments, is important for designing and developing corrosion prevention solutions. Protective coatings are often used to passivate the MIC effects of SRB on underlying metal surfaces, specifically to serve as a barrier against corrosive metabolites. Protective coatings are typically based on polymers [1], polymer composites [2], and inorganic materials including an emerging class based on atomic layers of two-dimensional materials such as graphene [3] and hexagonal boron nitride [4]. Here, our goal is to automate the extraction of the geometric properties (shape, size, etc.) of SRB cells from the scanning electron microscope (SEM) images of the SRB generated at various growth stages. Typically, these geometric features are extracted and measured using laborious and manual methods. To automate this process, we combine image processing and deep learning methods to extract the geometric properties automatically from the biofilm images.

Microscopy image analysis tools, including BiofilmQ [5], ImageJ [6], BioFilm Analyzer [7], and Imaris [8], have been used successfully for microscopy image feature extraction and for measuring geometric properties of cell features (e.g., length of a rod-shaped bacterial cell from images). While these tools perform well when the microscopy images are characterized by homogeneous and non-overlapping features, they are not necessarily optimized to tackle crowded features, e.g., overlapping bacterial cells. Microscopy images of biofilms often display heterogeneities related to the shape and size of bacterial cells, cell clusters, pores, and microbial debris. In the case of MIC, the heterogeneities multiply due to the existence of corrosion products. The existing image analysis tools are not necessarily designed to automate the extraction of these heterogeneous features. For instance, BiofilmQ performs well in segmentation and data visualization but lacks the capability of segmenting individual bacterial cells when present in a cluster [5]. ImageJ, a tool widely used in microscopy image analysis, also underperforms in separating individual bacterial cells from clusters, as will be seen later in the results section.

Here we demonstrate the ability of deep learning combined with image processing algorithms to extract the microscale geometric features of biofilms. Particularly, we develop (a) a mask region-convolutional neural network (Mask R-CNN) [9] based approach for instance segmentation of the bacterial cells (Figure 1); (b) a deep convolutional neural network model (using DeepLabV3+ [10]) combined with a modified watershed algorithm for semantic segmentation of the bacterial cells (Figure 2). These two models are integrated with the moment invariants approach to extract the geometric properties of the segmented cells. Next, we compare the performance of the above methods against a commercial microscopy tool ImageJ.

————————————————

*Mailing address for S. Ragi, H. Rahman, J. Duckworth: South Dakota Mines, 501 E. Saint Joseph St., Room EEP 310, Rapid City, SD 57701, USA.*
*Mailing address for K. Jawaharraj, V. Gadhamshetty: South Dakota Mines, 501 E. Saint Joseph St., CM 237B, Rapid City, SD 57701, USA.*
*Mailing address for P. Chundi: University of Nebraska Omaha, 1110 South 67th Street, Peter Kiewit Institute 281A, Omaha, NE 68182, USA.*

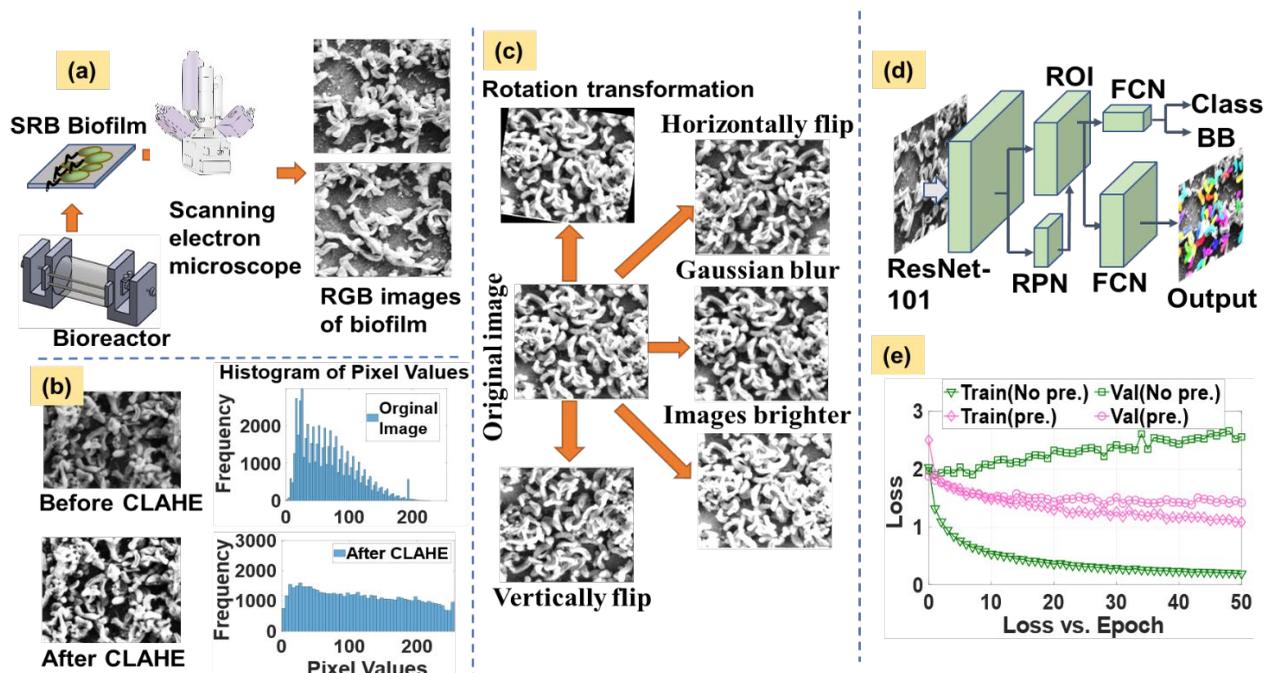

Figure 1. (a) SEM images generated at various growth stages of SRB biofilms; (b) Image pre-processing via CLAHE; (c) Data augmentation to increase data volume and avoid overfitting; (d) Mask R-CNN model for instance segmentation of bacterial cells; (e) Training and validation loss functions with and without the pre-processing step (CLAHE and data augmentation)

## 2 METHODS

Deep learning aided tools are being developed in application domains including medical image analysis [11-14], computer vision [15-17], speech recognition [18-20], self-driving cars [21, 22], object detection [23, 24], semantic segmentation [25-27], instance segmentation [28, 29], and image generation [30-32]. Deep learning is fundamentally composed of a deep artificial neural network structure with several layers that can learn the high-dimensional hierarchical features of objects from large training datasets using a backpropagation algorithm [33], that is typically used to train the network, while minimizing the error between the predicted and the actual labels. We implement two deep neural network architectures to segment bacterial cells in the SEM images and extract geometric features from the segmented images. In the first approach, we implement a deep convolutional neural network (DCNN) model using DeepLabv3+ (DLv3+) platform [10] (based on semantic segmentation), which is then integrated with a modified watershed algorithm [34] to segment both the individual and clustered bacterial cells. In the second approach, we implement another convolutional neural network model called Mask R-CNN [9], for instance segmentation of the bacterial cells. In both the approaches, we use the moment invariants approach [35] to automate the extraction of the geometric size properties of the segmented bacterial cells (area, length, width, and perimeter of the cells). Next, we benchmark the performance of our methods against a commercial image analysis tool called ImageJ. We also benchmark the performance of our methods against the ground-truth, which is manually measured by domain experts at the Two-Dimensional Materials for Biofilm Engineering, Science and Technology (2D-BEST) center, specifically the co-authors in studying MIC prevention on technologically relevant metals [2-4, 36].

### 2.1 DA-G20 Growth and Data Collection

Axenic cultures of DA-G20 were grown in Lactate C media as a carbon source with the following constituents in g/L: sodium lactate (6.8), dehydrated calcium chloride (0.06), sodium citrate (0.3), sodium sulfate (4.5), magnesium sulfate (2), ammonium chloride (1), potassium phosphate monobasic (0.5) and yeast extract (1). After inoculation, the sterile lactate media was deoxygenated using sterile $N_2$ gas for 20 min at 15 psi. Cultures were incubated at 30 °C under shaking conditions at 125 rpm for 48h and the exponential phase cultures were used for the MIC studies [32]. Mild steel samples coated with the biofilm were immersed in 3% glutaraldehyde in cacodylate buffer (0.1 M, pH 7.2) for 2 hours. The treated samples were rinsed with sodium cacodylate buffer and distilled water. The samples were then dried using ultra-pure nitrogen gas, followed by SEM imaging of the dried samples to characterize the SRB-G20 biofilm on mild steel surfaces (Figure 1(a)). We used 66 SEM images of the biofilm to train the deep learning models discussed next.

### 2.2 DeepLabv3+ Based DCNN

DLv3+ [28] was previously developed for semantic image segmentation, which uses atrous convolution [37] derived from a wavelet transform method a`trous ("hole algorithm" in French). A key component of this DCNN model is the use of an "encoder" that encodes the multiscale contextual information of the input image for a segmentation task by penetrating the incoming feature with atrous spatial pyramid pooling operations at different rates and feasible fields of view. Atrous convolution allows us to increase the filter's field of view by introducing holes into the filters and capturing features at multiple scales. A second key component 'decoder' then captures sharper object boundaries by gradually recovering the spatial information from the encoding phase and creates an output that is the size of the original input image.

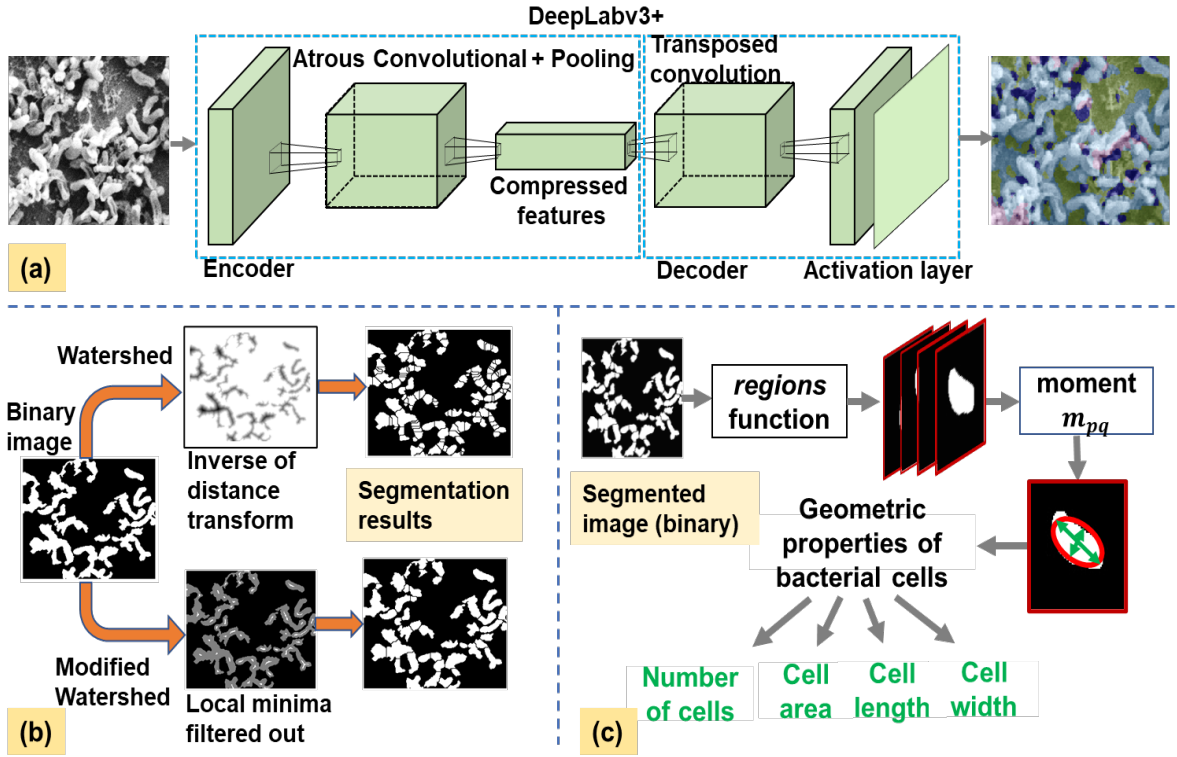

Figure 2. (a) Object segmentation in the biofilm SEM images via DeepLabv3+; (b) Segmentation process of the bacterial cell in a cluster via watershed and modified-watershed algorithms; (c) The extraction process of the geometric properties (area, length, width, and perimeter) of the bacterial cells in the biofilm image via the moment invariants approach.

Next, using a pixel annotation tool [38], we label the bacterial cells, pores, and other corrosion products (henceforth called CPs) in the training image datasets with guidance from domain experts at the 2D-BEST center [4]. We then use MATLAB's *pixelLabelDatastore* [39] object to read the labeled image data and store the pixel label data for semantic segmentation of the four object classes: cells, CP, pores, and the background surface. As shown in Figure 3, the model takes a raw unprocessed SEM image of a biofilm and outputs an image with segmented objects along with size characteristics of the bacterial cells present in the image.

Since our goal is to estimate the size characteristics of DA-G20 cells, we suppress the other object classes in the image by masking the pixels that belong to the other classes. Next, we determine if the image contains cell clusters via a certain criterion explained in Section 3.2. If clusters are present, we implement a modified watershed algorithm to separate individual bacterial cells in the cluster. Next, we apply feature extraction procedures to estimate the size properties of the bacterial cells. A detailed discussion of the feature extraction procedures is discussed in Section 3.2.

### 2.3 Mask R-CNN

Figure 1(d) shows the working principle of Mask R-CNN [9] in the context of the segmentation of our biofilm images. This model extends Faster R-CNN [40], a process that detects objects in an image and generates a segmentation mask for each object and is shown to outperform other object segmentation models [41]. Mask R-CNN is used in many applications such as segmenting nuclei in microscopy images [42], label-free cell tracking in phase-contrast microscopy [43], fruit detection for strawberry harvesting robot [44], characterization of arctic ice-wedge polygons in very high spatial resolution aerial imagery [44]. The basic principle of the model is to use a pre-trained backbone neural network, e.g., ResNet101 [45], that extracts the desired feature map of the objects in the image. These features would then be passed through both the region proposal network (RPN) and the region of interest alignment network (ROI Align). At this stage, the RPN layer returns the candidate bounding boxes and the ROI Align layer proposes candidate regions in the object's location. This is followed by a fully connected neural network, which then performs classification and bounding box regression on the ROI. Then, fully convolutional net (FCN) algorithm generates a binary mask for the targeted object. Each bounding box has at most one class and the FCN maps each input pixel to a '1' or '0', where '1' represents the presence of an object and '0' is the background.

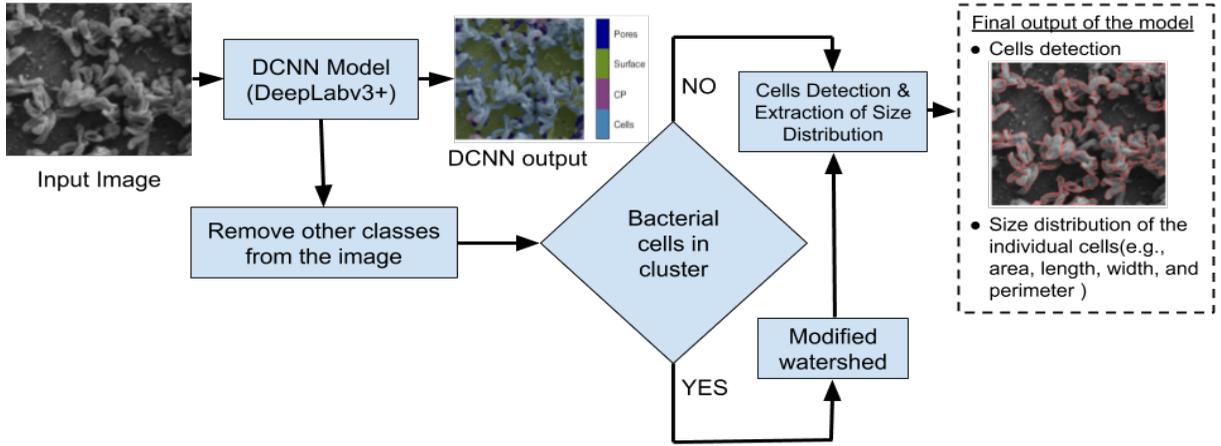

Figure 3. Feature extraction from biofilm SEM images via DCNN (DeepLabv3+) and watershed methods.

## 2.4 Bacterial Cell Geometry Extraction Approach

To count the number of bacterial cells, we label individual cells by numbers, as shown in Figures 4(b) (DeepLabv3+) and 4(c) (Mask R-CNN). To extract the geometric properties of the bacterial cells in a binary image (area, length, width, and perimeter of the individual cells), we use the moment invariants method [31] explained as follows. The moment $m_{pq}$ of oder (p + q), where p and q are non-negative integers, of a 2D image $f(x,y)$ is defined in [46, 47] as

$$m_{pq} = \sum_{x=1}^{w}\sum_{y=1}^{h} x^p y^q f(x,y), \quad (1)$$

where $f(x,y)$ is the pixel intensity of the image array at $(x,y)$, and $w$ and $h$ are the width and height of the image in pixels. In our case, the image is binary with the pixel intensity being either 0 or 1. The area of a segmented object (or cell) in the binary image is given by the zeroth moment, $m_{00}$ given by [46]

$$m_{00} = \sum_{x=1}^{w}\sum_{y=1}^{h} x^0 y^0 f(x,y) = \sum_{x=1}^{w}\sum_{y=1}^{h} f(x,y), \quad (2)$$

where $m_{00}$ is the area of the object.

The typical cell shape in the binary image is irregular, thus it is nontrivial to measure the length and the width of the object. Moment invariants allow us to infer an equivalent ellipse that best fits with the shape of the DA G20 cells. From Figure 4(a), we see that the length and the width of both the "fitted" ellipse and the cell are approximately equal. First, DLv3+ and Mask R-CNN (discussed above) are used to generate an individual mask for each bacterial cell. Then, we convert each image into a binary image, as seen in Figure 4(a). However, the DLv3+ model generates all the individual cells in a single binary image. To address this, we separate each bacterial cell into a separate polyshape object using MATLAB's 'regions' [48] function, which outputs an array of polyshape objects. Each element of this array represents a bacterial cell region. We then calculate the centroid $c$ and orientation $\theta$ of the object, and the semi-major axis $a$ and semi-minor axis $b$ of the ellipse via the following equations [46, 49],

$$c = \left(\frac{m_{10}}{m_{00}}, \frac{m_{01}}{m_{00}}\right) \quad (3)$$

$$\theta = \frac{1}{2}\tan^{-1}\frac{2m_{11}}{m_{20}-m_{02}} \quad (4)$$

$$a = \sqrt{\frac{m_{20}+m_{02}+\sqrt{(m_{20}-m_{02})^2+4m_{11}^2}}{\frac{m_{00}}{2}}} \quad (5)$$

$$b = \sqrt{\frac{m_{20}+m_{02}-\sqrt{(m_{20}-m_{02})^2+4m_{11}^2}}{\frac{m_{00}}{2}}} \quad (6)$$

where $m_{10}, m_{01}$, and $m_{11}$ are the first-order moments and $m_{02}$ and $m_{20}$ are the second-order moments obtained from Equation (1). Next, we calculate the perimeter, of the object in the binary image using the following expression [50],

$$2\pi\sqrt{\frac{a^2+b^2}{2}}. \quad (7)$$

## 3 RESULTS AND DISCUSSION

### 3.1 Key Findings

The execution time to extract the geometric shape of the bacterial cells of the Mask R-CNN and DLv3+ models are 227x and 70x, respectively, faster than manual measurement by the domain experts. Our cross-validation analysis confirms that the *f1-score* (from the cross-validation matrix) for Mask R-CNN is 1.06x higher than ImageJ and 1.03x higher than DLv3+. Finally, our numerical studies confirm that the Mask R-CNN model is significantly more accurate than ImageJ and DLv3+ in terms of measuring the geometric properties such as area, length, width, and the number of the bacterial cells in a biofilm microscopy image.

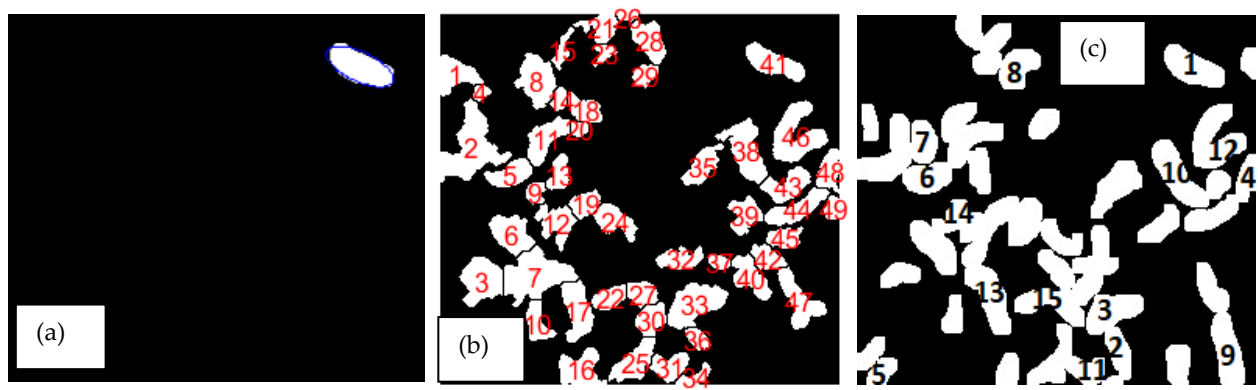

Figure 4. (a) A bacterial cell (white area) overlayed with an ellipse (blue color). (b) DLv3+ based automated feature extraction from biofilm SEM images. (c) Mask R-CNN based automated feature extraction from biofilm SEM images.

## 3.2 Image Preprocessing

The performance of a deep-learning neural network typically depends on the size and quality of the dataset and the variability in the experimental conditions to generate the dataset. Our dataset is limited to just 66 SEM images of the biofilm, and some of the images have "blurred" features (e.g., edges) as can be seen in Figure 1(b), which may degrade the cell detection and segmentation performance. To address these issues, we implement: a) Contrast Limited Adaptive Histogram Equalization (CLAHE) algorithm [51] for contrast enhancement; b) data augmentation methods [52] including rotation, horizontal/vertical flipping, multiplication, and Gaussian blurring transformations to increase the data volume, thereby avoiding overfitting of the model to the training dataset.

The histogram equalization from CLAHE enhances the contrast of an image by spreading the pixel histogram, thus reducing the "blurriness" in the object-background boundary. In this approach, an input image is divided into blocks and the histogram values of each block region are calculated. However, if there is noise in any block, the approach will amplify the noise that can lead to overfitting issues. To address this issue, the desired clip limit for clipping histograms is applied to each block of the image, and the resulting histogram of the image is redistributed in a way that never crosses the desired clip limit, effectively limiting the impact of the noise. Finally, the histogram equalizer redistributes the pixel intensity values of the input image more uniformly into new pixel intensity levels through a cumulative density function (CDF). We tune the CLAHE parameters "block size" and "contrast limit" via trial-and-error with different values to obtain the best contrast label of the image by observing the histogram of the redistributed pixel values. An image with "good" histogram (in the current context) is one where all the pixel intensity values have an equal number of pixels [53]. The results from the CLAHE algorithm on a sample SEM image is shown in Figure 1(b), with the SEM images before and after CLAHE, and their corresponding histograms. To increase the volume of the training dataset, we use data augmentation techniques [52] including rotation, horizontal/vertical flipping, multiplication, and Gaussian blurring transformations as shown in Figure 1(c). The horizontal and vertical flip operations generate new images, which are then added to the training data. We applied two rotation operations, where the first operation chooses a random value between -45 and +45 degrees, and the second operation between -90 and +90 degrees. A Gaussian blur transformation is also applied with *sigma* randomly chosen between 0 and 5.

## 3.3 Bacterial cell segmentation via deep learning methods

The parameters used in training DLv3+ are shown in Table 1. Adapting the learning rate of the model can increase the prediction performance and reduce the training time. We use the *'piecewise'* learning rate schedule option in DLv3+ that multiplies the learning rate by a factor of 0.3 every 10 epochs from the initial learning rate of 0.001 as this adaptive learning rate allows the model to learn relatively quickly. The '*ValidationData*' parameter is used for testing the validation data for every epoch to check for overfitting. To avoid the network from overfitting on the training dataset, '*ValidationPatience*' parameter is set to 4 to stop training early if the validation accuracy converges. We set the *'min-batch size'* parameter to 8, which means the network is trained by a batch of 8 images for each instance. The model continues this procedure until all the images from the training dataset are utilized.

Our image dataset has 66 images: 44 are used for training, 11 for validation, and 11 for testing. As with training any neural network, the validation images are used to tackle model overfitting, and the testing images are used for determining the model accuracy. We tested three backbone networks namely ResNet-18 [45], ResNet-50 [45], and MobileNet-V2 [54], to train the DLv3+ model with different iterations to assess the model's accuracy; the results from these iterations as shown in Table 2. We choose ResNet-50 as the backbone network to train our neural network upon since it achieves the highest model accuracy as can be seen in Table 2. We trained the DLv3+ model on a computer with an on-board GPU (NVIDIA GTX 16 Series, 6 GB memory) and used CUDA to enable

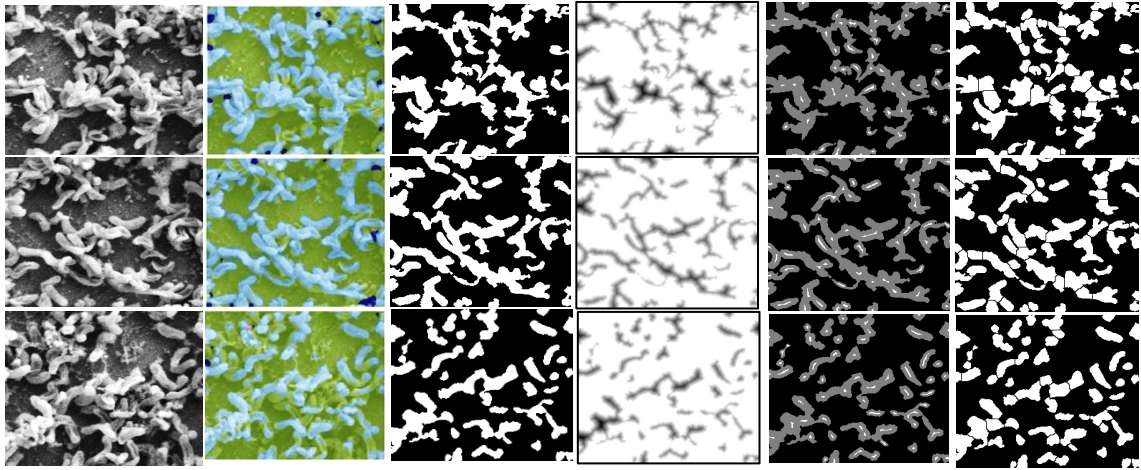

Figure 5. From left to right column: raw SEM images of DA G20 biofilm on mild steel surfaces; the outputs of the DeepLabv3+ model; converted into binary image; inverse of the distance transform; after filtering out the tiny minima, where grey areas as the bacteria group and lighter grey areas as the centers of clusters found from the modification; new segmentation applied from the modified watershed algorithm.

GPU-based acceleration to reduce to the training time.

The segmentation results from the trained DLv3+ model are shown in Figure 5, where the raw SEM image is shown in the first column, and the segmented cells are shown in the second column. The moment invariants method we use to extract the geometric properties of the bacterial cells requires the images to be in binary format, where '1' in a pixel represents bacterial cells and '0' represents the background. So, we convert the grayscale image into binary images, which are shown in the third column of Figure 5.

A key challenge with detecting bacterial cells in the biofilm images is separating the cells from the clusters. To address this, we implement a modified watershed transform method [34] to separate the individual cells from the clusters. First, we compute the distance transform of the masked clusters in the input images, which is achieved by allocating a number for each binary zero-valued pixel that represents the distance between the said pixel and the nearest non-zero pixel of the given image. Then, the modified watershed algorithm takes the inverse of these values as a way to weigh the values – this allows to find the minimum value in the region that should represent the center of a cell cluster. The fourth column of Figure 5 shows the inverse of the distance transform of a masked image [55]. A maxima edge finding algorithm is then used to find dividing sections between all logged minima [56]. We observe that this standard watershed algorithm leads to undesirable over-segmentation, as seen in the fifth column of Figure 5 as each small local minima becomes a "catchment basin". To address this issue, we implement a modified distance transform method [57] as explained below. First, MATLAB tool *imextendedmin* is applied to find the inverse of the distance transform for the image in the first column of Figure 5 to find local minima. This function helps the watershed segmentation filter-out the local minima in each segment which mitigates the over-segmentation issue by only returning the minimum value per segment. The resulting image has only one minima per segment, as can be seen in the fifth column of Figure 5. Then, with this new distance transformation method, we reapply the watershed transform for cell segmentation. The resulting cell segmentation is shown in the sixth column of Figure 5, which shows that the above modified watershed method overcomes the over-segmentation issue.

We use the same image datasets to train the Mask R-CNN model we used in training the DLv3+ model with the exception of using a different annotation tool, VGG Image Annotator (VIA) [58], to annotate the dataset for the Mask

TABLE 1
THE TRAINING PARAMETERS USED FOR DEEPLABV3+

| Parameter | Value |
|---|---|
| Initial learning rate | 0.001 |
| Maximum epoch | 200 |
| Learning rate schedule | Piecewise |
| Learning rate reduced by a factor | 0.3 every 10 epochs |
| Min-batch size | 8 |
| Maximum iterations | 1000 |

TABLE 2
MODEL ACCURACY AT DIFFERENT ITERATIONS

| Network | Model Accuracy at Different Iterations | | |
|---|---|---|---|
| | 150 | 500 | 1000 |
| ResNet-18 | 70.65% | 70.37% | 71.12% |
| ResNet-50 | 74.78% | 74.67% | 74.24% |
| MobileNet-v2 | 61.09% | 60.12% | 61.12% |

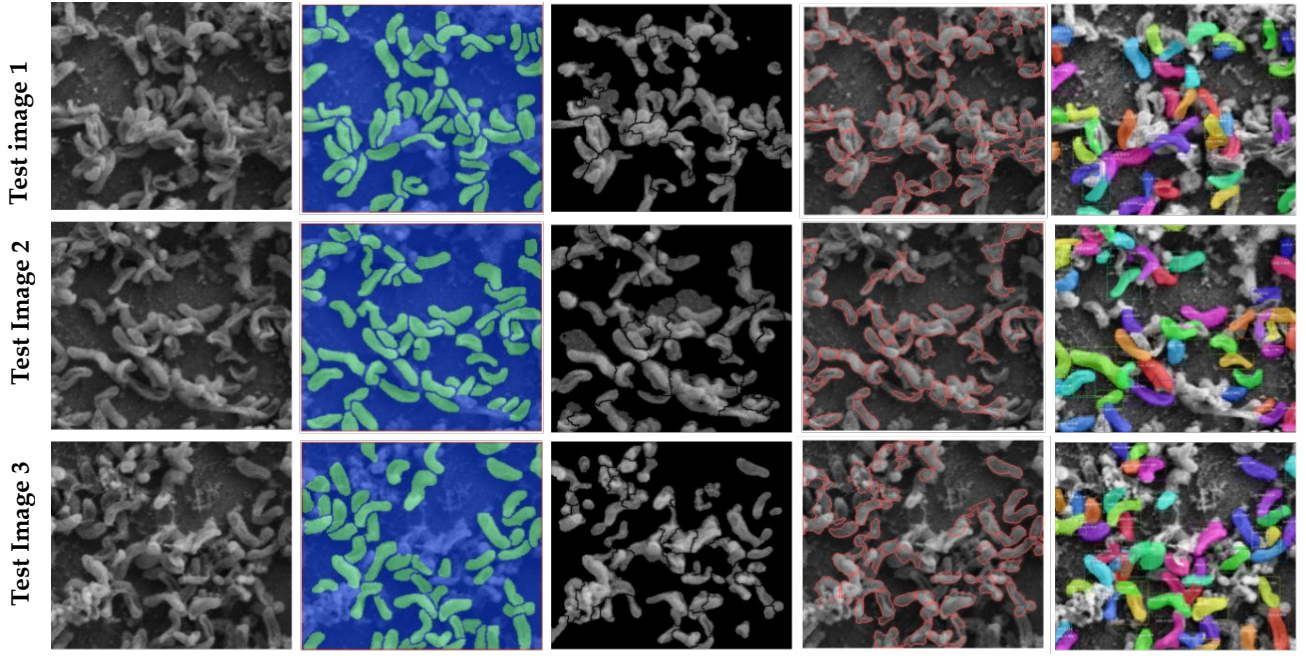

Figure 6. From left to right column: raw SEM images of biofilm, ground truth for cell segmentation (manually labelled), automatic segmentation from DLv3+, segmentation via ImageJ, and instance segmentation via Mask R-CNN. 'Test Image X' refers to the SEM images generated in identical experimental conditions, where X is the image index.

TABLE 3
THE TRAINING PARAMETERS USED FOR MASK R-CNN

| Parameter | Value |
|---|---|
| Learning momentum | 0.9 |
| Weight decay | 0.0001 |
| Learning rate | 0.001 |
| No. of epochs | 200 |
| Iterations per training epoch | 200 |

R-CNN training. This tool helps in generating compatible annotated files for MS-COCO [41], a pre-trained network whose weights we use to train Mask R-CNN instead of training the model from scratch for savings in training time. The parameters used for Mask R-CNN training are shown in Table 3. We trained the Mask R-CNN model on the same computer we used to train DLv3+ and reduced the training time via GPU-based acceleration using CUDA tools. The duration of the training process was 10 hours on average. Each image of the dataset is automatically resized to 1024 x 1024 px² where the model preserved the aspect ratio with any remaining space zero-padded. Next, we convert the annotations to JSON format compatible with the MS-COCO dataset. We then divide the annotated dataset into train, validation, and test datasets in the ratio of 3:1:1. We used a multitask loss function to train the Mask R-CNN model on the annotated dataset [9],

$$L = L_{cls} + L_{bbox} + L_{mask} \qquad (8)$$

where $L_{cls}, L_{bbox}, L_{mask}$ represent the classification loss, bounding-box regression loss, and the average binary cross-entropy loss, respectively.

The loss function curves of the training process are shown in Figure 1(e). The dotted curves represent the validation loss, and the solid curves represent the training loss. As discussed previously, we preprocessed the training dataset using the CLAHE approach. We trained the network with and without the image preprocessing step to quantitatively verify the impact of this preprocessing step on the loss functions. The blue curves indicate the loss without preprocessing, and the orange curves indicate the loss with the preprocessing step. Figure 1(e) suggests that the validation losses are significantly reduced with the preprocessing step.

The cell segmentation results from the two deep learning approaches discussed above are shown in Figure 6, along with the results from ImageJ (commercial microscopy image analysis tool) and the ground truth (manually labeled with help from domain experts at 2D-BEST center [4]). The size of a bacterial cell varies across growth conditions and thus the determination of their sizes is crucial. However, the mechanism of their differential cell size depends upon their differential gene expression resulting in a different phenotype. This genotypical and phenotypical changes are dependent on the microbial biofilm interaction at the materials-microbe interface, and they are relatively ambiguous.

To extract the size properties of the segmented bacterial cells (area, length, width, and perimeter of the individual cells), we use the moment of invariants method. Here, we used three test images of DA-G20 biofilms grown on mild steel surfaces from identical MIC experiments to validate the performance of all the methods discussed in this study.

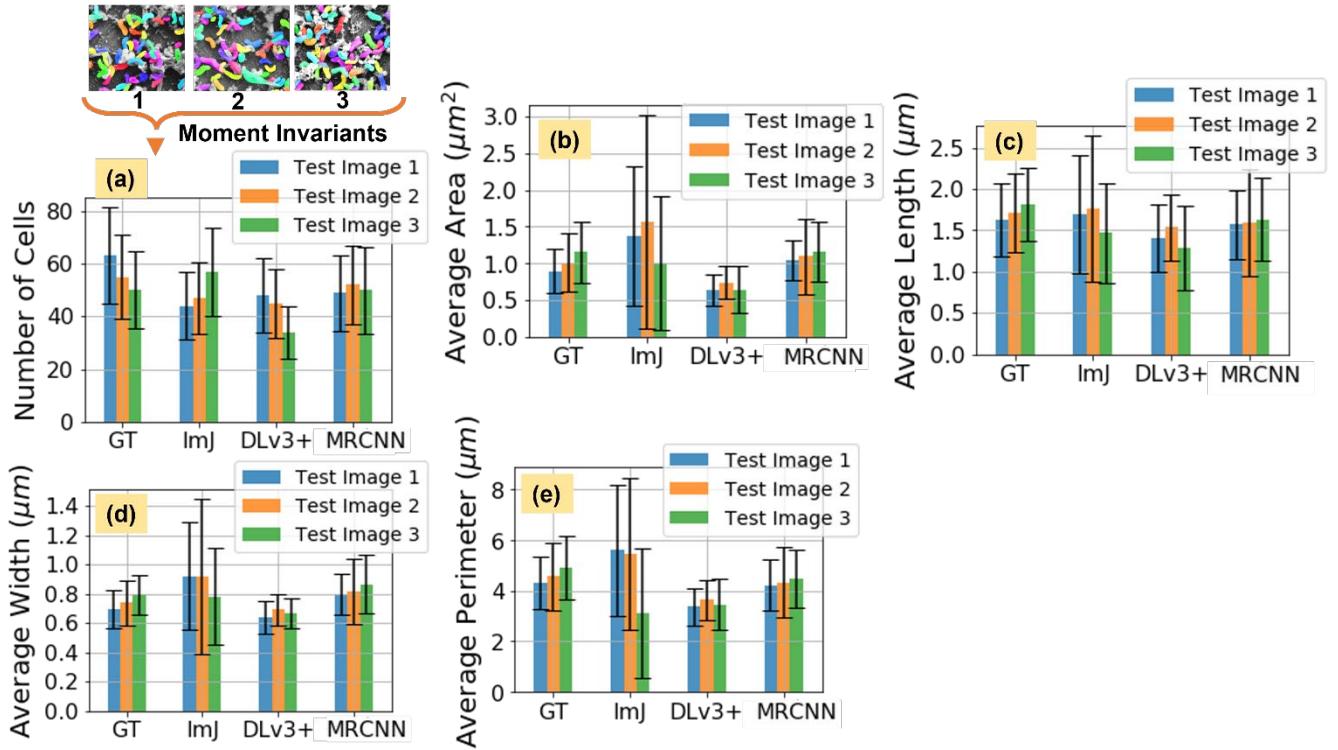

Figure 7. Geometric properties of the segmented cells estimated via different segmentation approaches (color bars represents row of Figure 4). (a) number of cells (b) average area of the cells (c) average length of the cells (d) average width of the cells (e) average perimeter of the cells. The error bars represent 'one' standard deviation of the mean. Acronyms GT, ImJ, DLv3+, and MRCNN represent ground truth, ImageJ, DeepLabV3+, and Mask R-CNN respectively. Here 'Test Image X' refers to the biofilm SEM images generated in identical experimental conditions, where X is the image index.

TABLE 4
ESTIMATED GEOMETRIC PROPERTIES OF BACTERIAL CELLS

| Model | | No. of cells | Avg. area ($\mu m^2$) | Avg. length ($\mu m$) | Avg. width ($\mu m$) | Avg. perimeter ($\mu m$) |
|---|---|---|---|---|---|---|
| T1 | GT | 63 | 0.89 | 1.62 | 0.69 | 4.32 |
|    | ImJ | 44 | 1.36 | 1.69 | 0.92 | 5.59 |
|    | DLv3+ | 48 | 0.63 | 1.40 | 0.64 | 3.35 |
|    | MCNN | 49 | 0.95 | 1.56 | 0.75 | 4.31 |
| T2 | GT | 55 | 1.01 | 1.70 | 0.74 | 4.57 |
|    | ImJ | 47 | 1.56 | 1.76 | 0.92 | 5.46 |
|    | DLv3+ | 45 | 0.74 | 1.53 | 0.69 | 3.64 |
|    | MCNN | 52 | 1.15 | 1.62 | 0.76 | 4.23 |
| T3 | GT | 50 | 1.15 | 1.81 | 0.80 | 4.91 |
|    | ImJ | 57 | 1.01 | 1.46 | 0.78 | 4.52 |
|    | DLv3+ | 34 | 0.64 | 1.28 | 0.67 | 3.11 |
|    | MCNN | 57 | 1.10 | 1.52 | 0.81 | 4.24 |

*T1, T2, and T3 represent the Test Image 1, Test Image 2, and Test Image 3, respectively. GT= Ground Truth, ImJ=ImageJ, MCNN= Mask R-CNN*

These images are referred to as Test Images 1, 2, and 3. The geometric feature extraction results are shown in Table 4. Figure 7 shows the average values and the corresponding error bars for the estimated geometric properties over all the segmented cells. Figures 7(b), 7(c), 7(d), and 7(e) suggest that the DLv3+ model suffers from the significant spread in estimating the area, the length, the width, and the perimeter of the bacterial cells, which is not observed in the ground truth measurements. This clearly suggests DLv3+ may not perform as well as the other approaches including Mask R-CNN and ImageJ. In counting the number of cells, both DLv3+ and Mask R-CNN methods perform close to the ground truth as seen in Figure 7(a). With respect to the overall estimation of the cell geometric properties and counting the number of cells, Mask R-CNN outperforms both DLv3+ and ImageJ, as is evident from Figure 7.

It is crucial to establish how confident a trained model is in forecasting data that has not been seen yet. Here, we use the Dice similarity coefficient ($F_1 - Score$) to evaluate the model's segmentation performance of the bacterial cells. $F_1 - Score$ is used for measuring the model's classification accuracy on the test dataset. First, we evaluate the precision and the recall values of each test image for all methods as follows [19]:

$$precision = \frac{TP}{TP + FP} \quad (10)$$

$$recall = \frac{TP}{TP + FN} \quad (11)$$

where True Positive (TP) is defined as the number of correctly segmented bacterial cells, False Negative (FN) is the number of unsegmented bacterial cells, and False Positive (FP) is the number of incorrectly segmented bacterial cells. The TP, FN, and FP values are computed using Python programming and averaged over each test image. Next, we evaluate the $F_1 - $ Score using equations (10) and (11) as follows:

$$F_1 - score = \frac{2 \times precision \times recall}{precision + recall} \quad (12)$$

The corresponding performance of each method applied to the test images is presented in Table 5, which confirms that Mask R-CNN clearly outperforms the other methods considered in this study.

TABLE 5
SEGMENTATION EFFICIENCY EVALUATION PERFORMANCE

| Methods | Precision | Recall | $F_1 - score$ |
|---|---|---|---|
| ImageJ | 0.73 | 0.73 | 0.73 |
| DeepLabv3+ | 0.75 | 0.75 | 0.75 |
| Mask R-CNN | 0.81 | 0.74 | 0.77 |

Furthermore, we measure the execution times for each method discussed above, which includes the time needed for cell segmentation and the extraction of the size properties of the bacterial cells. We also evaluate the execution times for the manual measurement approach (cells identified and the geometric properties measured by domain experts at 2D-BEST center) and the ImageJ tool. We use 'timeElapsed' function for DeepLabv3+ and 'timeit' library for Mask R-CNN to measure the execution times. Table 6 summarizes the findings in terms of the time taken to complete each experiment, where the execution times are averaged over all the cells in each of the three test images considered. From Table 6, we conclude that DeepLabv3+ and Mask R-CNN models outperform both the manual approach and the analysis via ImageJ. Specifically, DeepLabv3+ and Mask R-CNN models are 70x and 227x faster, respectively, than the manual measurement approach by the domain experts. In summary, the Mask R-CNN approach is found to be the best choice (among the methods considered here) in terms

TABLE 6
THE EXECUTION TIME OF EACH METHOD FOR THE TEST IMAGES (T1, T2, AND T3 REPRESENT THE TEST IMAGES 1, 2, AND 3 RESPECTIVELY)

| Image | Manual | ImageJ | DLv3+ | Mask R-CNN |
|---|---|---|---|---|
| T1 | 23m 34s | 3m 26s | 18.97s | 5.68s |
| T2 | 20m 15s | 3m 10s | 17.87s | 5.50s |
| T3 | 18m 54s | 3m 06s | 17.08s | 5.48s |
| Average | 20m 54s | 3m 14s | 17.97s | 5.53s |

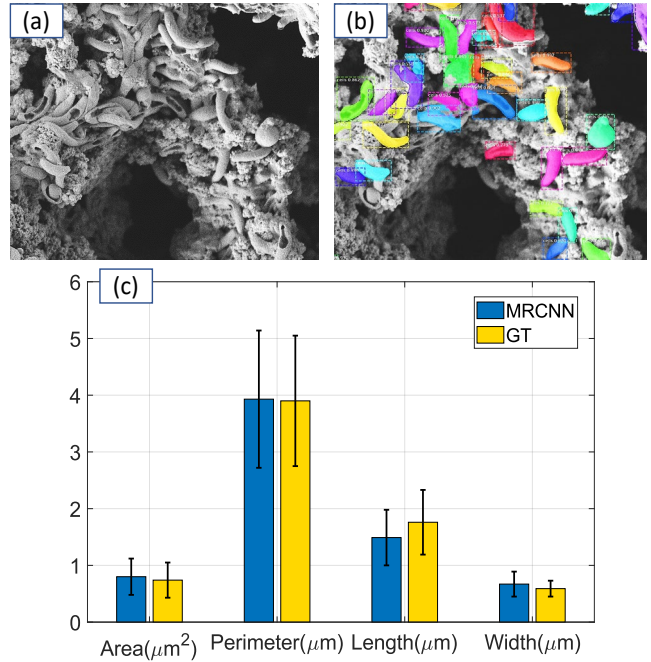

Figure 8. Performance of Mask R-CNN and *moment invariants* for cell segmentation and cell size estimation on a biofilm system - DA-G20 formed on 56.2% cold-worked copper. (a) Raw SEM image of the biofilm after 70 days of growth; (b) Instance segmentation of bacterial cells from Mask R-CNN approach; (c) Estimation of size properties from - the *moment invariants* approach applied on the segmented images from Mask R-CNN and the ground truth (error bars represent one standard deviation).

of the segmentation accuracy and the execution time.

To assess the robustness of the Mask R-CNN model, we test the model's performance for cell segmentation and cell size estimation on a new microbial corrosion system where DA-G20 cells are grown on copper substrates (images from this system are not used in the training phase). Specifically, we grew DA-G20 cells on 56.2% cold-worked copper samples exposed to corrosion cells. These tests lasted for 70 days, and the samples were also incubated for the same time. The raw SEM images, cell segmentation results from the Mask R-CNN approach, and the cell size estimation results from the moment invariants approach are shown in Figures 8(a), 8(b), and 8(c) respectively. The results in Figure 8(c) demonstrate that Mask R-CNN performs close to that of the ground truth in size property estimation of the bacterial cells. It should be noted that the biofilm system discussed in the previous sections is DA-G20 cells grown on mild steel. In summary, the results shown in Figure 8 validate the performance of Mask R-CNN (not retrained) on biofilm systems grown on diverse metal substrates for cell segmentation and cell size property estimation. The diverse metal systems represent diverse metal infrastructure, which stands as a proof-of-concept to the proposition that these deep learning methods can be extended to other bacterial systems for structural phenotyping.

## 4 CONCLUSIONS

We developed two deep learning-based image segmentation approaches to automate the extraction of geometric size properties of bacterial cells in biofilms, which is otherwise a laborious and time-consuming process. Particularly, we studied sulfate-reducing bacteria-based biofilms, which are widely implicated in microbial-induced corrosion of metals. Rapid screening and selection of protective coatings against MIC effects of biofilms require high-throughput microscopy image characterization methods to automate the process of measuring structural changes in the biofilms in response to the coatings. To aid this process, we implemented two neural network architectures based on DeepLabv3+ and Mask R-CNN for the segmentation of bacterial cells in the biofilm microscopy images. Furthermore, we implemented the moment invariants method to extract the size properties of the segmented bacterial cells from the biofilm images. Our numerical study confirmed that Mask R-CNN outperforms both DeepLabV3+ and ImageJ methods in terms of estimation accuracy of the geometric properties of the bacterial cells in the biofilm and demonstrated that the Mask R-CNN and DLv3+ models are 227x and 70x, respectively, faster (execution time) than manual measurement by the domain experts.

Although the deep learning models presented here can be extended to MIC studies on other metal surfaces, heterogeneities on the biofilm surface may pose a significant challenge. For instance, previous studies confirm that with DA-G20 grown on copper surfaces, the biofilm produces complex mesh-like microstructures in the EPS matrix, which may degrade the performance of the cell segmentation methods discussed in this study. To address this issue, one may first need to "filter" out these microstructures in the image pre-processing stage before they are passed to the deep learning models. Without re-training, our methods can also be readily extended to other biofilm systems including those that of certain pathogens (e.g., *E. Coli),* bacteria responsible for pitting corrosion (e.g., *P. aeruginosa*) and biotechnologically relevant bacteria (*B. subtilis*) where the shape and the structure of the bacterial cells are similar to that of the DA-G20 biofilm. With minimal or no retraining, our models can be extended to other biotechnologically relevant biofilm systems, even when the shape and the structure of the bacterial cells are significantly different from that of DA-G20. The deep learning models presented in this study (i.e., the codes that implement the models) are accessible to the scientific community through GitHub at https://github.com/hafizur-r/BiofilmScanner-v0.1.

## ACKNOWLEDGMENTS

The authors acknowledge funding support from NSF RII T-2 FEC award #1920954. S. Ragi would like to acknowledge NSF RII T-1 FEC award #1849206 for a seed grant that partially supported this study. V. Gadhamshetty would like to acknowledge partial support from NSF CAREER award #1454102. Dr. Gadhamshetty's group is thankful to Dr. Bharat Jasthi, Materials and Metallurgical Engineering (MET), SD Mines for providing copper samples for the dislocation experiments.

## DATA AVAILABILITY

The data that support the findings of this study are available from the corresponding author upon request.

## CODE AVAILABILITY

The source code, the trained network weights, and the training data are available at https://github.com/hafizur-r/BiofilmScanner-v0.1

## CORRESPONDING AUTHOR

Shankarachary Ragi – Department of Electrical Engineering, South Dakota Mines, 501 E. Saint Joseph St., Room EEP 310, Rapid City, SD 57701, U.S.A., orcid.org/0000-0002-5511-4334; Phone: 605-391-3218; Email: Shankarachary.ragi@sdsmt.edu

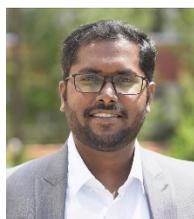
**Shankarachary Ragi** is an assistant professor in the Electrical Engineering Department at South Dakota Mines, USA. He earned his Ph.D. degree in Electrical and Computer Engineering at Colorado State University, USA in 2014, and his B.Tech. and M.Tech. degrees in Electrical Engineering at Indian Institute of Technology Madras, India in 2009. Before joining South Dakota Mines, Ragi has worked as a postdoctoral researcher in the mathematics department at Arizona State University, and prior to that, he worked as a Senior Controls Engineer at Cummins Emission Solutions. He is currently serving as a senior personnel at the 2D-Materials for Biofilm Engineering, Science and Technology (2D-BEST) center funded by the National Science Foundation. His current research interests include machine learning, image analysis, robotics, and optimal control. Ragi has served as an Associate Editor for IEEE Access during 2017-2020. He has authored or co-authored over 26 peer-reviewed publications in various journals and conference proceedings. He is a senior member of the IEEE.

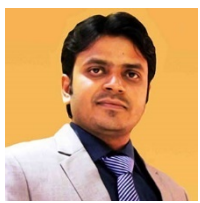
**Md Hafizur Rahman** has completed his M.S. in Electrical Engineering at South Dakota Mines and his B.Sc. in Electrical and Electronic Engineering at Pabna University of Science and Technology, Bangladesh. During his M.S., he worked on data-driven models for biofilm phenotype prediction on metal surfaces modified with 2D coatings.

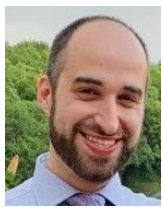
**Jamison Duckworth** is currently working toward his M.S. thesis in Electrical Engineering at South Dakota Mines. He holds a B.S. degree in Physics with a minor in Mathematics from Creighton University. He previously worked as a researcher at University of Nebraska Medical Center and Creighton University. His current research areas include machine learning, 2D image analysis, and fractal mathematics.

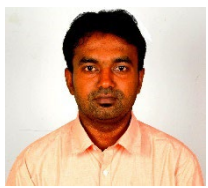
**Kalimuthu Jawaharraj** is currently working as a Research Scientist III at the Civil and Environmental Engineering Department, South Dakota Mines for the past three years. He has completed MS in Biotechnology from SRM University, India and a Ph.D. in Biotechnology from Madurai Kamaraj University, India. During his Ph.D., he worked on the genetic modification of cyanobacteria for biodiesel production. His current research at SD Mines involves biofilm engineering, biofuels from methanotrophs and wastewater treatment infrastructures.

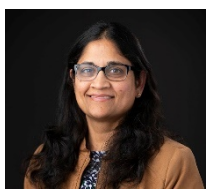
**Parvathi Chundi** is a professor in the Computer Science Department at the University of Nebraska-Omaha. She is a University Distinguished Professor and the director of the Big Data and Text Mining Lab. Her research interests include data management, machine learning, and text mining. She has published over 80 research articles in peer-reviewed conferences and journals and has been a primary investigator on several research projects in machine learning and text mining. She worked at several industrial research labs including HP Labs and Agilent Labs prior to joining the academia. Currently, she is working on applying machine learning methods for extracting knowledge from text and image collections related to the 2D materials and biofilms domains.

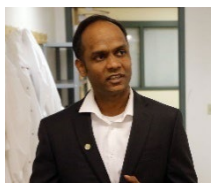
**Venkataramana Gadhamshetty** earned a PhD degree in Civil and Environmental Engineering from New Mexico State University, MS degree in Environmental Engineering from National University of Singapore, and BS degree in Chemical Technology from Osmania University. **He is currently** an Associate Professor in Civil and Environmental Engineering department at South Dakota Mines, USA. He has over a decade of teaching and research experience from South Dakota Mines, Rensselaer Polytechnic Institute, Florida Gulf Coast University, Air Force Research Laboratory and industrial experience from Dupont Singapore Pte Ltd. He is a Board-Certified Environmental Engineer, a licensed Professional Engineer, and the chair of the ASCE EWRI Water Pollution Engineering Committee. He is a recipient of the National Science Foundation CAREER award (2015), South Dakota Mines Research Award (2016), and an invited Tedxtalk speaker for Rapid City in 2017. His research on bioelectrochemistry were featured by BBC, CNN, American Chemical Society, History Now, and 350 other large media outlets. He is a thrust area lead and core investigator (co-I) at 2D-Materials for Biofilm Engineering, Science and Technology (2D-BEST) center and for other projects funded NSF, NASA EPSCoR, and Electric Power Research Institute. He has served as investigator or senior personnel for projects worth ~$32 MM. His ongoing projects interrogates the fundamental phenomena at the interface of 2D materials and biofilms. Examples of practical outcomes from these projects include NASA microbial fuel cells and infinitesimally thin coatings for corrosion applications.